\documentclass[aps,prl,twocolumn,groupedaddress]{revtex4-2}

\usepackage{float}
\usepackage{graphicx}
\usepackage{amsmath}
\usepackage{relsize}
\usepackage{hyperref}

\begin{document}

\title{Comment on ``Distinct Behaviors of Inner and Outer CuO$_2$ Planes in Quadruple-Layer Cuprate
(Cu,C)Ba$_2$Ca$_3$Cu$_4$O$_{11 + \delta}$"}

\author{J. L. Tallon}

\affiliation{Robinson Research Institute, Victoria University of Wellington,
P.O. Box 33436, Lower Hutt 5046, New Zealand.}



\noindent {\bf Comment on ``Distinct Behaviors of Inner and Outer CuO$_2$ Planes in Quadruple-Layer Cuprate
(Cu,C)Ba$_2$Ca$_3$Cu$_4$O$_{11 + \delta}$"}

In a recent Letter, Sun {\it et al.} \cite{Sun} report photoemission spectroscopy measurements on the four-layer cuprate (Cu,C)Ba$_2$Ca$_3$Cu$_4$O$_{11 + \delta}$ (CuC-1234) in which they resolve two superconducting gaps associated with the inner (IP) and outer (OP) CuO$_2$ planes. CuC-1234 has a high $T_{\textrm{c}}$ of 110 K in the sample investigated and as high as 117 K in a sample previously structurally characterized using neutron powder diffraction \cite{Shimakawa}. From the Luttinger sum, the (degenerate) $\beta$-band of the IP was found to be heavily underdoped ($p\approx0.07$) while the OP $\alpha_2$ bonding band was strongly overdoped ($p\approx0.25$). Each revealed a gap with very different momentum and temperature dependences. A large gap on the $\beta$ band was found to close at the bulk $T_{\textrm{c}}$, while a smaller gap on the $\alpha_2$ band closed at $75 \pm 5$ K. Such two-gap behavior would indicate weak-coupling between the IP and OP, resulting in a second gap-opening temperature, $T_{\textrm{c2}}$, and implying a two-step development of the superfluid density on cooling, first on the IPs then on the OPs. The $\alpha_1$ antibonding band was even more overdoped ($p\approx0.32$) and this remained ungapped, as expected, as it lies outside the dome.

Here, we point out that the observed values of $T_{\textrm{c}}$ and $T_{\textrm{c2}}$ are fully consistent with a long-standing correlation of $T_{\textrm{c}}$ with a bond-valence sum (BVS) parameter, $V_+$, calculated from the crystallographic bond lengths \cite{Tallon1990,Mallett}. The parameter $V_+ = 6-V_{\textrm{Cu}}-V_{\textrm{O2}}-V_{\textrm{O3}}$, comprising the in-plane copper and oxygen BVS, characterizes how much the hole content resides on oxygen orbitals relative to copper orbitals and is reproduced in Fig. 1 from ref. \cite{Mallett}. $T_{\textrm{c}}^{\textrm{max}}$ is the maximum $T_{\textrm{c}}$ on the $T_{\textrm{c}}(p)$ doping dome. For more detail, including the effect of planar stretch or contraction (red or blue crosses, respectively) see \cite{Mallett}. 
The correlation is excellent.

From the crystallographic data for (Cu,C)1234 \cite{Shimakawa} we calculate $V_+$ for both the IP and OP and plot $T_{\textrm{c}}$ and $T_{\textrm{c2}}$ by the red squares. We use $T_{\textrm{c}}$ = 117 K as for the sample characterized in ref. \cite{Shimakawa}. The two data points match the correlation well. And if we take the value of $T_{\textrm{c2}}$ = 80 K based on the loss of the coherence peak (Fig. 4(g) in Sun {\it et al.} \cite{Sun}) then we have an almost perfect fit (upper error bar for the OP data point). This suggests that $T_{\textrm{c2}}$ in the OP is entirely consistent with the local structure of the OP where the value of $V_+$(OP) is reduced by the presence of the apical oxygens and $T_{\textrm{c2}}$ reduced accordingly. It is as though two largely decoupled superconductors are present in the single 1234 compound and {\it their individual pairing temperatures are determined by their local structure} reflecting the very short-range physics.  

Yet it is more complex than this because {\it both the IP and OP are each behaving as if optimally doped}. In fact the IP is heavily underdoped, and the pseudogap and associated `Fermi arcs' are present in the relevant $\beta$ band \cite{Sun} as expected, yet $T_{\textrm{c}}$ is not reduced below $T_{\textrm{c}}^{\textrm{max}}$ as it would be in an underdoped bilayer cuprate where the depleted spectral weight severely reduces both $T_{\textrm{c}}$ and the superfluid density \cite{Bernhard,Tallon2026}. We feel that the idea of adjacent OP conducting layers suppressing phase fluctuations in the IP, as was suggested \cite{Sun}, cannot be the full answer as, in the case of YBa$_2$Cu$_3$O$_7$ the adjacent conducting chains enhance the superfluid density with little effect on $T_{\textrm{c}}$ \cite{TallonChains}. Either the IP must pick up compensating spectral weight from the OP or the pseudogap plays a rather different (and weaker) role in CuO$_2$ planes which lack an apical oxygen. Perhaps it is the former because there is a similar problem with the strongly-overdoped OP $\alpha_2$ band where $T_{\textrm{c2}}$ (and its associated gap) should also be severely reduced below its own $T_{\textrm{c}}^{\textrm{max}}$ \cite{Bernhard}. It is not. Meanwhile, the lack of a gap on the $\alpha_1$ band (lying outside the dome) remains entirely as expected. These ideas should also be applicable to other reported $n$-layer cuprates, with $3\leq n\leq 9$ \cite{Wang}, and in particular $V_+$ will likely correlate with the charge transfer gap for each layer. 
\linebreak
\linebreak
\noindent Jeffery L. Tallon

\noindent Robinson Research Institute, Victoria University of Wellington, New Zealand; jeff.tallon@vuw.ac.nz

\begin{figure}[H]
\centering
\includegraphics[width=0.85\columnwidth]{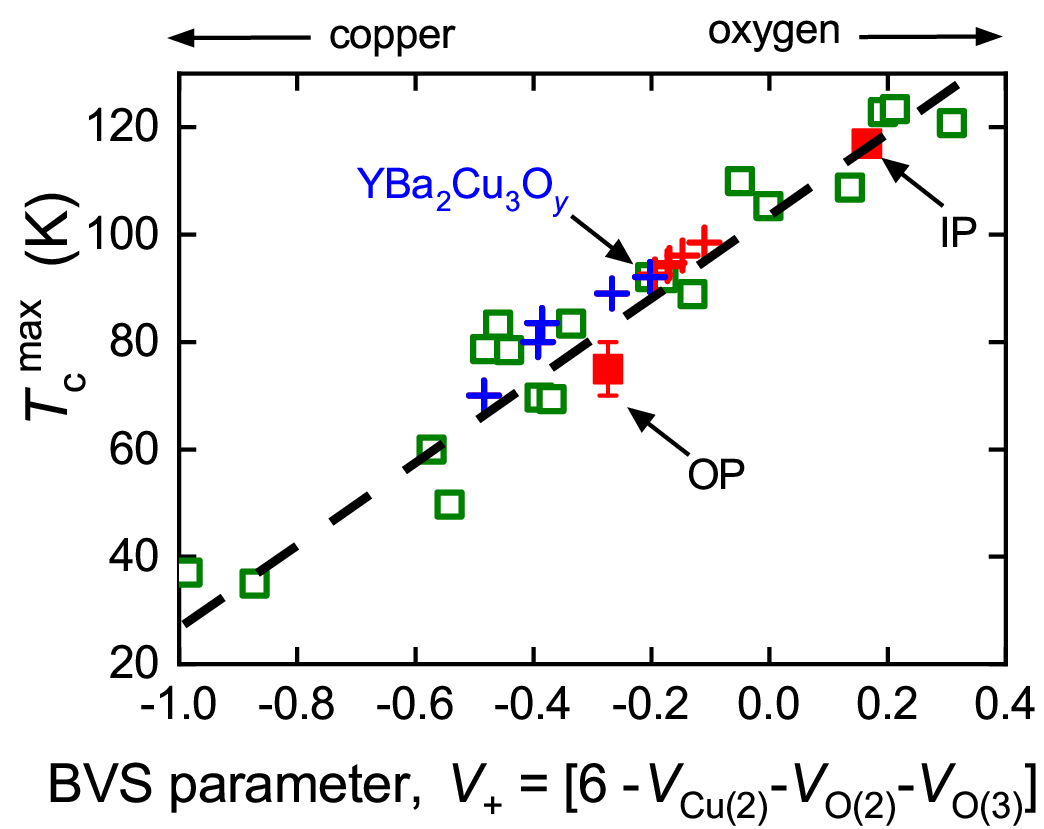}
\label{fig:bvs}
\caption{$T_{c,\textnormal{max}}$ plotted vs $V_+ = 6 - V_{\textnormal{Cu}} - V_{\textnormal{O}2} - V_{\textnormal{O}3}$ after refs. \cite{Tallon1990,Mallett}. Olive squares: cuprates as first reported \cite{Tallon1990}; red crosses: RBa$_2$Cu$_3$O$_y$ (R = La, Nd, Sm, Gd, Dy, Yb); blue crosses: YBa$_{2-x}$Sr$_x$Cu$_3$O$_y$ (x = 0, 0.5, 1.0, 1.25, 2) \cite{Mallett}. Full red squares: $T_{\textrm{c}}$ and $T_{\textrm{c2}}$ versus $V_+$(IP) and $V_+$(OP), respectively, for CuC-1234.  }
\end{figure}

\end{document}